# Observation of dynamic non-Hermitian skin effects

Zhen Li[1,*], Li-Wei Wang[2,*], Xulong Wang[1], Zhi-Kang Lin[2], Guancong Ma[1,†], Jian-Hua Jiang[2,3,†]

[1]Department of Physics, Hong Kong Baptist University, Kowloon Tong, Hong Kong, China

[2] School of Physical Science and Technology, and Collaborative Innovation Center of Suzhou Nano Science and Technology, Soochow University, 1 Shizi Street, Suzhou 215006, China

[3]Suzhou Institute for Advanced Research, University of Science and Technology of China, Suzhou 215123, China

[*]These authors contributed equally to this work.

[†]Correspondence should be addressed to: phgcma@hkbu.edu.hk (GM) or jianhuajiang@suda.edu.cn (JHJ)

**Non-Hermitian effects have emerged as a new paradigm for the manipulation of phases of matter that profoundly changes our understanding of non-equilibrium systems, introducing novel concepts such as exceptional points and spectral topology, as well as exotic phenomena such as non-Hermitian skin effects (NHSEs). Most existing studies, however, focus on non-Hermitian eigenstates, whereas dynamic properties of non-Hermitian systems have been discussed only very recently, predicting unexpected phenomena such as wave self-healing, chiral Zener tunneling, and the dynamic NHSEs that are not yet confirmed in experiments. Here, we report the first experimental observation of rich non-Hermitian skin dynamics using tunable one-dimensional nonreciprocal double-chain mechanical systems with glide-time symmetry. Remarkably, dynamic NHSEs are observed with various dynamic behaviors in different dynamic phases, revealing the intriguing nature of these phases that can be understood via the generalized Brillouin zone and the related concepts. Moreover, the observed tunable non-Hermitian skin dynamics and amplifications, the bulk unidirectional wave propagation, and the boundary wave trapping provide promising ways to guide, trap, and amplify waves in a controllable and robust way. Our findings unveil the fundamental aspects and open a new pathway toward non-Hermitian dynamics, which will fertilize the**

**study of non-equilibrium phases of matter and give rise to novel applications in information processing.**

In the past decades, non-Hermitian systems, parity-time symmetry[1], and related effects led to surges of research interest and new discoveries in electron systems[2,3], classical systems[4–7], and open quantum systems[8–11]. In particular, the concept of exceptional points and their generalizations that are featured with eigenstates coalesce and spectral degeneracy, have substantially deepened our understanding of dynamics without energy conservation, giving rise to unprecedented phenomena such as unidirectional invisibility[12] and spectral topology[13–15]. Recently, it was found that non-Hermitian effects in lattice systems can fundamentally modify the Brillouin zone---a cornerstone concept in condensed matter physics---leading to the notions of generalized Brillouin zones (GBZs)[16] and NHSEs[17–25]. These effects significantly change the properties of energy spectra and eigenstates, giving rise to non-Hermitian topology, which is drastically distinct from its Hermitian counterpart[26–33].

To date, most works on non-Hermitian systems focus on static eigenstate properties. However, eigenstates alone can be insufficient in the study of non-Hermitian dynamics, which is essential for understanding the properties of non-Hermitian systems. In general, it becomes challenging to understand the dynamics when the spectrum is complex, which can give rise to intricate temporal dependence in the amplitudes of eigenmodes. Moreover, in non-Hermitian systems, boundary and defect scatterings can bring about exotic phenomena, such as wave self-healing[34] and inelastic boundary scatterings[35] that are beyond the conventional understanding of dynamics. Recent theoretical predictions of unconventional non-Hermitian dynamics such as chiral Zener tunneling[36], anharmonic Rabi oscillations[37], non-Bloch dynamics[38–40], dynamic NHSEs[35,41,42], non-Hermitian edge burst[43,44], and wave self-acceleration[45,46] reveal that non-Hermitian dynamics can result in phenomena beyond the static properties of eigenstates. One of the key challenges in this emerging field is how to characterize and classify non-Hermitian dynamics where the study of dynamic phases and their transitions play key roles.

Here, we demonstrate rich non-Hermitian skin dynamics and non-Hermitian dynamic phases emerging in one-dimensional (1D) nonreciprocal double-chain mechanical systems with the glide-time reversal ($GT$) symmetry. In our design, two shifted Su-Schrieffer-Heeger (SSH) chains of coupled mechanical oscillators are connected with each other, where only the inter-chain hoppings are nonreciprocal. In our mechanical systems, the couplings are largely tunable, making the system suitable for studying non-Hermitian dynamics in a wide range of parameters to explore various non-Hermitian dynamic phases. With such a platform, we measure the non-Hermitian dynamics and classify them into different types. We invent new methods based on the time evolution in the GBZs and in the eigenmodes at the open boundary condition (OBC) to analyze and understand the versatile non-Hermitian dynamics. We further explore the dynamic phase transitions and reveal the underlying mechanisms via the OBC eigenstates.

Several remarks are in order. First, unlike the previously studied harmonic[4–7] and adiabatic[15] wave dynamics, which involve only small frequency ranges, the phenomena here are triggered by a short pulse spanning a wide frequency range such that the full-spectrum dynamic properties are revealed. Second, in our system, the nonreciprocal hoppings are perpendicular to, instead of parallel with, the wave propagation directions, which is a feature distinct from the previous studies[35,41,42,45,46]. Lastly, we remark that the effects studied in this work need no time-dependent manipulation, which is thus distinguished from both the space-time varying systems[47] and the transient NHSE that utilizes time-dependent complex-frequency excitation[48].

**Theory and nonreciprocal mechanical systems**

We start by considering three distinct types of dynamics in finite 1D systems, as schematically depicted in Fig. 1. The first type is the Hermitian dynamics, wherein an initial wave packet at the center evolves into two wave packets that propagate in opposite directions until they are reflected back again and again by the boundaries, in addition to a third wave packet that stays at the initial position (Figs. 1a, b). Eventually, these wave packets spread and diffuse into the entire system due to the dispersion. As there is no loss or gain, the wave energy is conserved in the whole process. The second type can be called the conserved non-Hermitian skin dynamics. Here, the non-Hermitian dynamics is featured with the directionally biased wave propagation in the bulk and the inelastic scattering at the boundaries (i.e., the wave reflected by the boundary has a different group velocity as compared with the incident wave)[35]. Eventually, all wave energy is accumulated at one of the edge boundaries (Fig. 1d). In this regime, the spectrum is real despite its non-Hermitian nature. Thus, the total wave energy in the system is conserved. The third type is the unconserved non-Hermitian skin dynamics which is featured with unidirectional wave propagation due to the NHSE (Fig. 1e) and exponential growth or decay of the total energy due to the complex spectrum (Fig. 1f). Both the conserved and unconserved non-Hermitian skin dynamics are generally called the dynamic NHSEs.

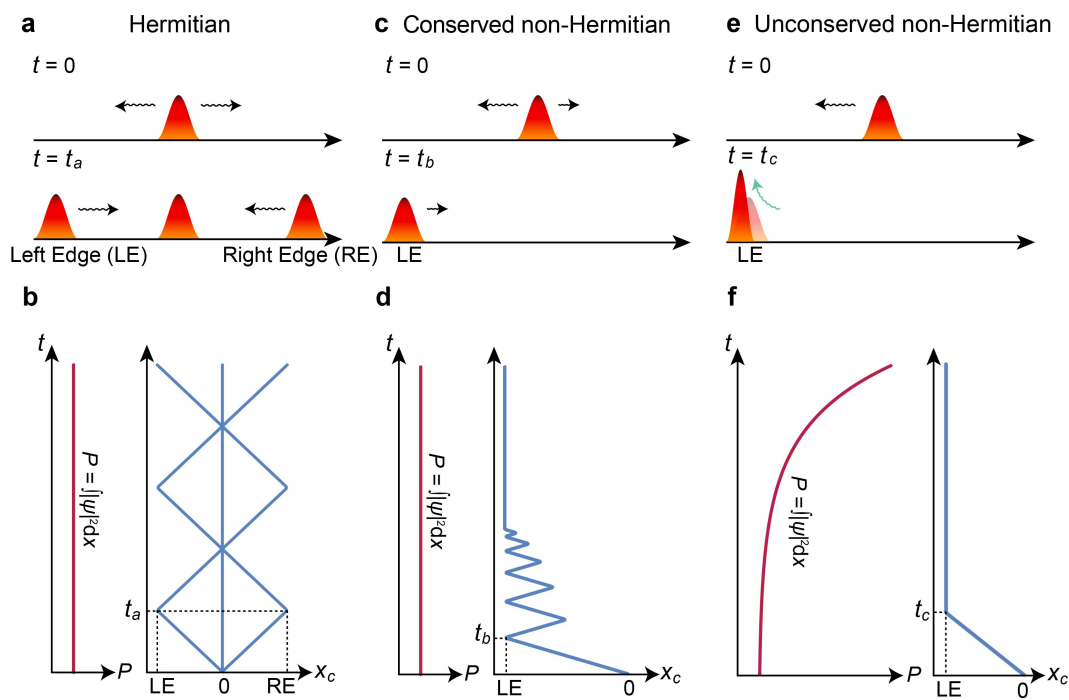

**Figure 1 | Schematic of three types of dynamics in 1D systems. a**, In the Hermitian regime, the initial wave packet at the system center splits into three wave packets: two wave packets propagate symmetrically to the left and right directions, a third wave packet is stationary. **b**, The two propagating wave packets are reflected by the left and right edge boundaries (as indicated by the central position of the wave packets $x_c$). The total wave energy $P$ is constant during the whole process. **c**, For the conserved non-Hermitian skin dynamics, the initial wave packet propagates biasedly towards one boundary. **d** The biased wave propagation in the bulk and the inelastic scattering at the boundary. The reflected wave packet decays in the bulk, so the wave gradually becomes localized at the left boundary, resulting in the dynamic NHSE. In this regime, the spectrum is real despite its non-Hermitian nature, such that the total wave energy is nearly conserved (it may have some nonexponential variations due to non-othorgonality of the non-Hermitian eigenstates). **e-f**, For the unconserved non-Hermitian skin dynamics, the initial wave packet propagates unidirectionally in the bulk and becomes localized at one of the edge boundaries. The complex spectrum gives rise to the exponential increase of wave energy. **d** and **f** also imply weak and strong dynamic NHSEs, respectively.

Next, we reveal various dynamic phases and dynamic phenomena in our system. The principle is illustrated by the dynamic phase diagram for various 1D non-Hermitian models in Fig. 2. Our goal is to show that as enriched by the crystalline symmetry, the non-Hermitian $GT$ model has a much richer dynamic phase diagram compared to the well-studied models like the Hatano-Nelson model and non-Hermitian SSH model. The non-Hermitian dynamics in the Hatano-Nelson model (Fig. 2a) is simple: When $t_1 > t_2$ ($t_1 < t_2$), the dynamic NHSE is toward the left (right) boundary, giving two dynamic phases I and II (Fig. 2b). This phase transition is stably protected by the time-reversal symmetry of the system. Similarly, the non-Hermitian SSH model, with both time-reversal symmetry and chiral symmetry, (Fig. 2c) also has only two dynamic phases (Fig. 2d): phases I and II with the same dynamic phase boundary ($t_1 = t_2$) as in the Hatano-Nelson model. Although there is a gap closing line (the dashed green line) in the phase diagram that separates the topologically and trivially gapped phases, it is not a dynamic phase boundary because the two gapped phases have the same dynamic properties: They are gapped in the spectrum and having conserved dynamic NHSEs (i.e., their spectra at the OBC are real).

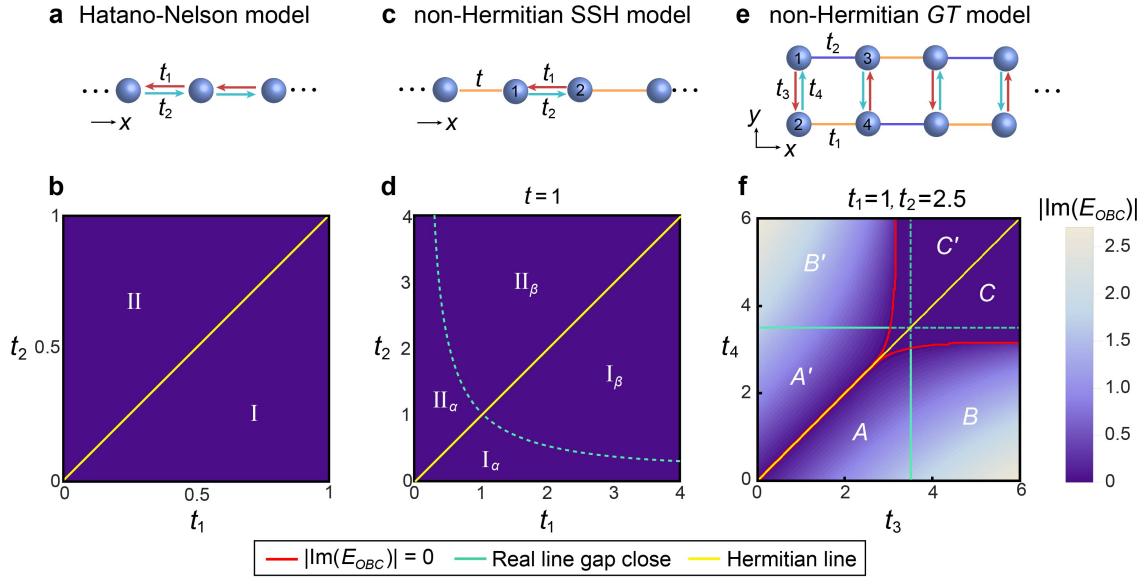

**Figure 2 | Dynamic phase diagrams for various non-Hermitian models. a**, The Hatano-Nelson model with nonreciprocal hoppings. **b**, The dynamic phase diagram of the Hatano-Nelson model. **c-d**, The non-Hermitian SSH model and its dynamic phase diagram. In **b** and **d**, phase I (II) has the dynamic NHSE toward the left (right) boundary. For the phases in **d**, the subscript $\alpha$ ($\beta$) labels the topologically (trivially) gapped phase. **e-f**, The non-Hermitian $GT$ model and its dynamic phase diagram. Yellow lines separate the phases with opposite dynamic NHSE directions. Green lines label the real line-gap-closing boundary. Red curves represent the phase boundary with $\text{Im}(E_{OBC}) = 0$. Dashed lines are marginal phase

boundaries, whereas solid lines label the true phase boundaries that separate phases with different dynamic behaviors. In all cases the OBC spectra are calculated for systems with 25 unit cells. The color scheme in **b**, **d**, and **f** represents the amplitude of $|\text{Im}(E_{OBC})|$. All hoppings are real and positive.

The non-Hermitian $GT$ model (Fig. 2e) has a much richer dynamic phase diagram. As shown in Fig. 2f, the Hermitian line, $t_3 = t_4$, separates the phase diagram into three pairs of phases. Phases $C$ and $C'$ feature an entirely real OBC spectrum, giving rise to the conserved dynamic NHSE. In contrast, in phases $A$, $A'$ and $B$, $B'$, the system's OBC spectrum has non-vanishing imaginary parts, leading to the unconserved dynamic NHSE. Furthermore, phases $A$ ($A'$) and $B$ ($B'$) are different in that $A$ and $A'$ have a real line gap under OBC, whereas $B$ and $B'$ are gapless. Besides, the difference between the unprimed ($A$, $B$, and $C$) and primed ($A'$, $B'$, and $C'$) phases is in the direction of the dynamic NHSE, which is controlled by the relative strength of $t_3$ and $t_4$: If $t_3 > t_4$ ($t_3 < t_4$), the dynamic NHSE is biased toward the left (right) boundary when $t_2 > t_1$. We will show that these dynamic phases exhibit distinct dynamic behaviors.

The Bloch Hamiltonian of the non-Hermitian $GT$ model is written as

$$H(k) = \begin{pmatrix} 0 & t_4 & t_2 + t_1 e^{-ik} & 0 \\ t_3 & 0 & 0 & t_1 + t_2 e^{-ik} \\ t_2 + t_1 e^{ik} & 0 & 0 & t_3 \\ 0 & t_1 + t_2 e^{ik} & t_4 & 0 \end{pmatrix}. \qquad (1)$$

Here, the system has two key symmetries: the time-reversal symmetry $T$ and the glide reflection symmetry $G$. The $G$ operation is a flip between the two SSH chains combined with a half-lattice-constant translation along the $x$ direction, i.e., $G := (x, y) \to (x + \frac{1}{2}, -y)$ (see Fig. 2e). There are several features unique to our model, as elaborated with great details in Supplementary Secs. I-III. Most importantly, the system does not have the parity-time symmetry, but instead the $GT$ symmetry. As a consequence of the $GT$ symmetry, NHSEs can be induced by the nonreciprocal hoppings that are perpendicular to the wave propagation directions. Empowered by this special setup, our system possesses versatile properties, e.g., real line gap closing and opening, topological phase transition, and two opposite directions of non-Hermitian dynamics, (see Supplementary Sec. I) which lead to rich dynamic phases and phase transitions. For instance, we find the three types of dynamics in the non-Hermitian $GT$ model, as shown in Figs. 3a-c: the Hermitian dynamics at $t_3 = t_4$, the conserved dynamic NHSE in phase $C$, and the unconserved dynamic NHSE in phase $B$. Moreover, the dynamic phase diagram of the model can be further tuned by the ratio $t_2/t_1$, as shown in Fig. 3d. We also find that in phases $A$ and $A'$, although there exist two topological edge modes in the bulk band gap, these edge modes do not play a notable role in the dynamics studied in this work. We thus ignore discussions on the topological properties of the system.

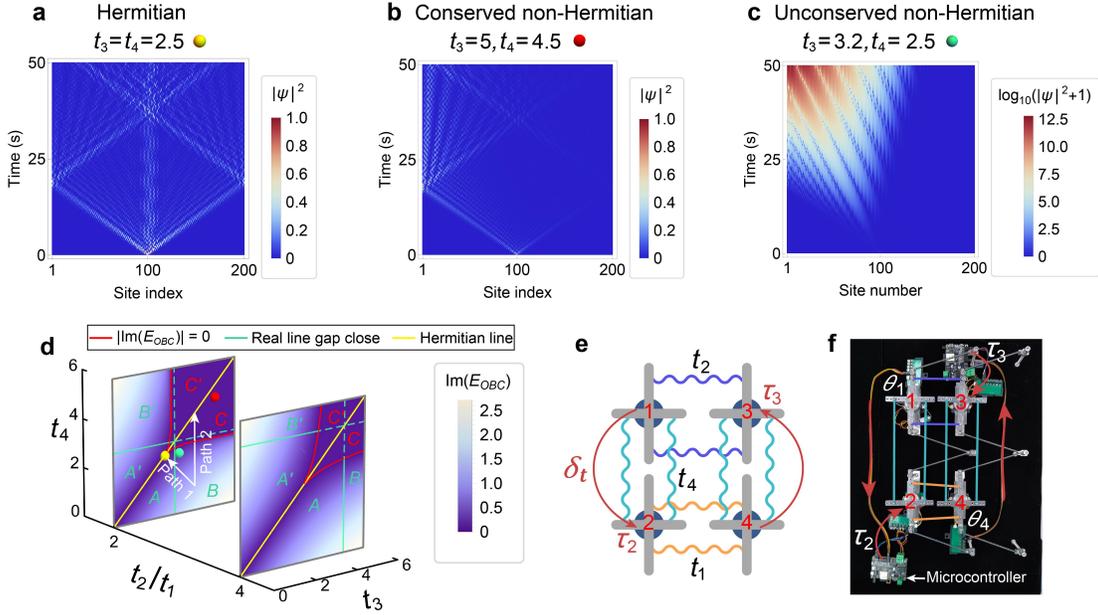

**Figure 3 | Non-Hermitian $GT$ model and its dynamics. a-c**, Three different types of dynamics in the model: Hermitian (**a**) and non-Hermitian (**b** and **c**). The horizontal axis represents the site index from left to right. Here, $t_1 = 1$, $t_2 = 2$, $t_3$ and $t_4$ are labeled on top of each figure. Conserved and unconserved non-Hermitian skin dynamics are observed in **b** and **c**, respectively (note that the color scheme in **c** is in the log scale). **d**, A phase diagram of the dynamics in an open-boundary lattice with 100 sites with the color scheme representing the amplitude of $|\text{Im}(E_{OBC})|$. The parameters are $t_1 = 1$ and $t_2 = 2, 4$. Red curves represent the phase boundary with $\text{Im}(E_{OBC}) = 0$. Green lines mark the real line-gap-closing positions in the OBC spectra. Yellow lines mark the Hermitian cases, i.e., $t_3 = t_4$. Yellow, red, and green dots label separately the parameters in **a**, **b** and **c**. White arrows indicate two paths for dynamic phase transitions that will be studied in Fig. 5. **e**, The schematic of a unit cell of the mechanical lattice that realizes the non-Hermitian $GT$ model. The gray crosses represent the rotational arms of the four mechanical oscillators. The orange, blue, and cyan curves (red arrows) represent reciprocal (nonreciprocal) springs corresponding to

the hoppings $t_{1,2,4}$ ($t_3 = t_4 + \delta_t$), respectively. **f**, Photograph of a unit cell of the mechanical lattice of phase *A*.

The non-Hermitian *GT* model can be realized in coupled mechanical oscillator systems, as illustrated in Figs. 3e-f (see Supplementary Fig. S9 for the full lattices). The two SSH chains are realized by mechanical oscillators coupled with tensioned springs. The nonreciprocal inter-chain hoppings are achieved by a special feedback design with programmed external actuation of the motors (Fig. 3f). The whole system consists of 10 unit cells (i.e., 40 mechanical oscillators). When the system is in the unconserved non-Hermitian regime, the mechanical oscillations may exhibit amplification if the positive imaginary parts of the spectrum exceed the intrinsic loss in each oscillator. This will result in the runaway breakdown of the experimental system. To prevent such breakdown, each oscillator can be attached in tandem with an additional motor that plays the role of an attenuator. These attenuators are connected to a tunable resistance to form a close circuit, such that the excess kinetic energy is dissipated by the Ohmic loss. This design allows us to realize tunable, uniform damping, which in turn can be used to quantify the amplification in the non-Hermitian dynamics of the original model. Such damping can be modeled as an additional constant negative bias in the imaginary part in all onsite terms of the Hamiltonian in Eq. (1). We have benchmarked the intrinsic damping of the mechanical oscillators without the attenuators to be $\gamma_0 \approx$ 2.64 rad/s or about 0.42 Hz. With the attenuators attached, the total damping is in the range of 0.49 to 1.06 Hz. We remark that the above design enables us to realize and observe all dynamic phases in the phase diagram (Fig. 3d) by tuning the hopping parameters and the damping (see Methods and Supplementary Sec. IV for details of the experimental setup).

**Experimental validation**

We measure the non-Hermitian dynamics for phases *A*, *B*, and *C*. The results are shown in Fig. 4. In all results, the excitation is a poke at site 20 with a constant rotation angle. In Fig. 4a, the spatiotemporal evolution of the mechanical oscillation is presented for a set of parameters in phase *A* (see the caption of Fig. 4 for the parameters). The unidirectional tendency of wave propagation is clearly observed, which confirms the dynamic NHSE. Because of the presence of homogeneous loss in the lattice, the imaginary parts of the eigenfrequencies are still negative. This enables us to observe the inelastic scattering from the boundary, i.e., the gradually reduced multiple bouncing back from the left edge boundary, as shown in Fig. 4a. These features are well reproduced by the calculation based on the same parameters as in the experiment (Fig. 4b). Careful examination of the dynamics at site 1 (one of the leftmost sites) indicates that there is a clear beating pattern at the frequency of 3.3 Hz (insets of Figs. 4a-b). This phenomenon originates from the coexistence of two modes with their real eigenfrequencies separated by the line gap illustrated in Fig. 4c. These two modes have the largest imaginary parts in the eigenfrequencies (corresponding to the smallest damping) such that they dominate the long-time dynamics. This scenario is also confirmed by performing a short-time Fourier transformation (STFT) of the measured oscillation dynamics at site 1 (Fig. 4d; see more details in Supplementary Sec. IV).

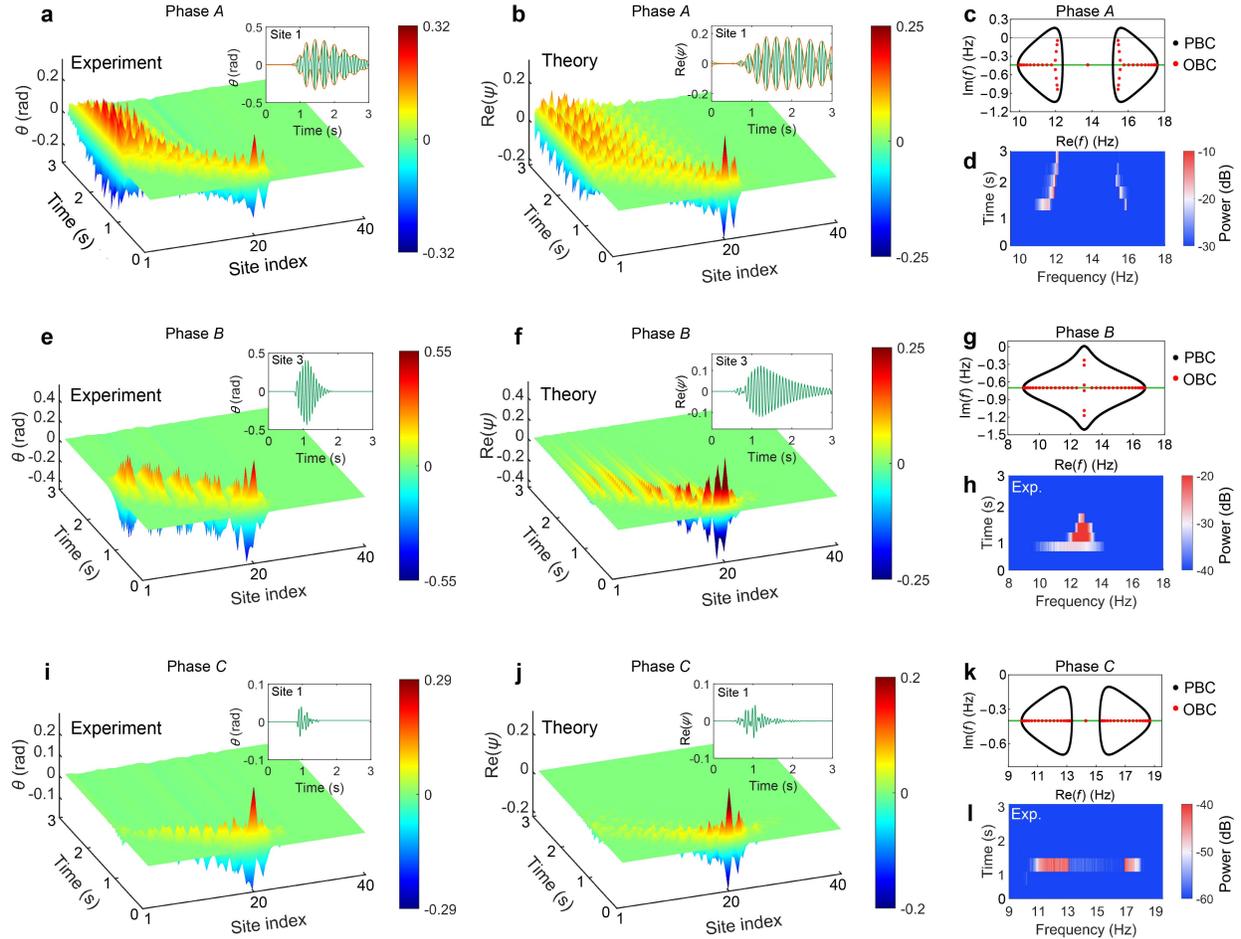

**Figure 4 | Non-Hermitian dynamics in different dynamic phases. a-b**, The measured instantaneous rotation angle $\theta$ and the calculated wavefunction $\text{Re}(\Psi)$ with a set of parameters in phase $A$ ($t_1 = 2.1$ rad/s, $t_2 = 14.9$ rad/s, $t_3 = 11.2$ rad/s, $t_4 = 3.7$ rad/s). One axis represents the site index numerating from the

left to the right of the 1D system. Here, $\omega_0 = 2\pi f_0 = 86.5$ rad/s and $\gamma = 2.8$ rad/s are the frequency and damping of the mechanical oscillators, respectively. The insets plot the data at site 1 with the envelopes marked by the orange curves. **c**, The corresponding PBC and OBC spectra from calculations. **d**, The STFT of the experimental data at site 1. **e-h**, The experimental and calculated results for a set of parameters in phase $B$ ($t_1 = 3.2$ rad/s, $t_2 = 6.7$ rad/s, $t_3 = 22.6$ rad/s, $t_4 = 8.4$ rad/s, $\omega_0 = 80.9$ rad/s, and $\gamma = 4.4$ rad/s). **i-l**, The experimental and calculated results for a set of parameters in phase $C$ ($t_1 = 2.1$ rad/s, $t_2 = 14.9$ rad/s, $t_3 = 12.6$ rad/s, $t_4 = 8.9$ rad/s, $\omega_0 = 89.8$ rad/s, and $\gamma = 2.5$ rad/s). In **c**, **g**, **k**, the green horizontal line denotes the net damping at each mechanical oscillator.

The spatiotemporal evolution of the mechanical oscillation is measured and presented in Fig. 4e for a set of parameters in phase $B$. The observed non-Hermitian dynamics agree well with the calculation based on the model with the same parameters (see Fig. 4f). Here, the OBC spectrum is gapless. There are several modes (corresponding to the frequency $f_0 = \frac{\omega_0}{2\pi} = 12.9$ Hz in the mechanical system) have large imaginary parts, as shown in Fig. 4g. The OBC spectrum also indicates that a damping $\gtrsim 0.6$ Hz must be used to suppress the amplification. Thus, the attenuators are employed to increase the total damping to 0.7 Hz. This fact alone already indicates unconserved non-Hermitian skin dynamics. Under this condition, the measured and calculated wave dynamics show strong boundary wave trapping (Figs. 4e-f), indicating strong dynamic NHSE. In addition, the beating pattern observed in phase $A$ is absent, which is due to the fact that phase $B$ is gapless. Here, a few modes with the largest imaginary eigenfrequency (see the OBC spectrum in Fig. 4g) dominate the dynamics soon after the excitation. Consequently, the observed dynamics at site 1 soon show typical single-frequency oscillation behavior with damping (insets of Figs. 4e-f) due to the designed attenuators. The STFT of the measured oscillation at site 1 also indicates such a scenario, even though in the beginning many other modes are excited as well (Fig. 4h).

We now turn to phase $C$, where the OBC spectrum is entirely real (in the absence of loss). Figure 4k shows the OBC spectrum biased by the intrinsic loss of 0.4 Hz. As shown in Fig. 4i, the spatiotemporal evolution here shows no beating pattern in the wave dynamics but unidirectional wave propagation in the bulk, which is clear evidence of the NHSEs. These features are theoretically confirmed as well (Fig. 4j). These observations conform well with the fact that phase $C$ is a conserved phase with negligible amplification. The STFT of the dynamic response at site 1 (Fig. 4l) gives two spectral bands separated by a spectral gap which corresponds to the line-gapped OBC spectrum shown in Fig. 4k. Here, due to the purely real spectrum and the short-pulsed excitation, many modes are involved in the dynamics. The complex interference among these modes destroys any beating pattern. We remark that because of the presence of the homogeneous loss bias, the wave packet significantly decays before reaching the edge. As a result, the inelastic scattering was unclear in Fig. 4i.

The non-Hermitian dynamics observed here is determined by the coaction of the NHSE and the complex energy of the OBC eigenmodes---both are related to the GBZs. To better illustrate the underlying physics, we perform the following analyses, which are presented in Fig. 5. First, we Laplace-transform the measured spatial wavefunctions at different times and project them to the generalized wavevector $\beta$ that forms the GBZs (Fig. 5a). It has been revealed that GBZs offer a useful framework to understand the non-Hermitian dynamics in finite systems[20,49]. Because of the $GT$ symmetry in our system, the spectrum is always symmetric about $\text{Re}(E) = 0$, and therefore, for each $\beta$ there are two different modes with opposite $\text{Re}(E)$. In addition, due to the enriched unit-cell structure, there are always two GBZs that touch at a point with $\text{Im}\beta = 0$ and $\text{Re}\beta < 0$ – a feature guaranteed by the $GT$ symmetry and holds for all parameters (see Supplementary Sec. II). We emphasize that symmetry plays a pivotal role in the structure and properties of the GBZs, which are crucial for understanding the non-Hermitian dynamics.

From the Laplace transformation results for phase $A$ in Fig. 5b, one finds that the condition at time $t = 0$ triggers most of the $\beta$ components in the GBZs due to the highly-localized and short-pulsed excitation in experiments. As time progresses, the wavefunction evolves towards modes with small $|\beta|$. This is clear evidence of the skin effect taking over the dynamic evolution – because here, the GBZ is entirely inside the BZ (the unit circle), smaller $|\beta|$ indicates a stronger skin effect towards the left boundary. The time evolution in phases $B$ and $C$ also display similar behavior, as shown in Figs. 5c-d. Second, to reveal the role of the imaginary part of the eigenfrequencies in the dynamics, we decompose the time-domain wavefunctions to the OBC eigenmodes (Fig. 5a). The results for the three cases are shown in Figs. 5e-g. In Figs. 5e-f, the evolution in the OBC spectrum converges to the modes with the largest imaginary eigenfrequencies (i.e., the modes having the smallest damping) in the long-time dynamics. This conforms well with the experimental results shown in Figs. 4a-h. For the case shown in Fig. 5g, where all OBC eigenfrequencies are real, the evolution converges towards the modes near the line-gap edges, which are also the modes with small $|\beta|$ in Fig. 5d.

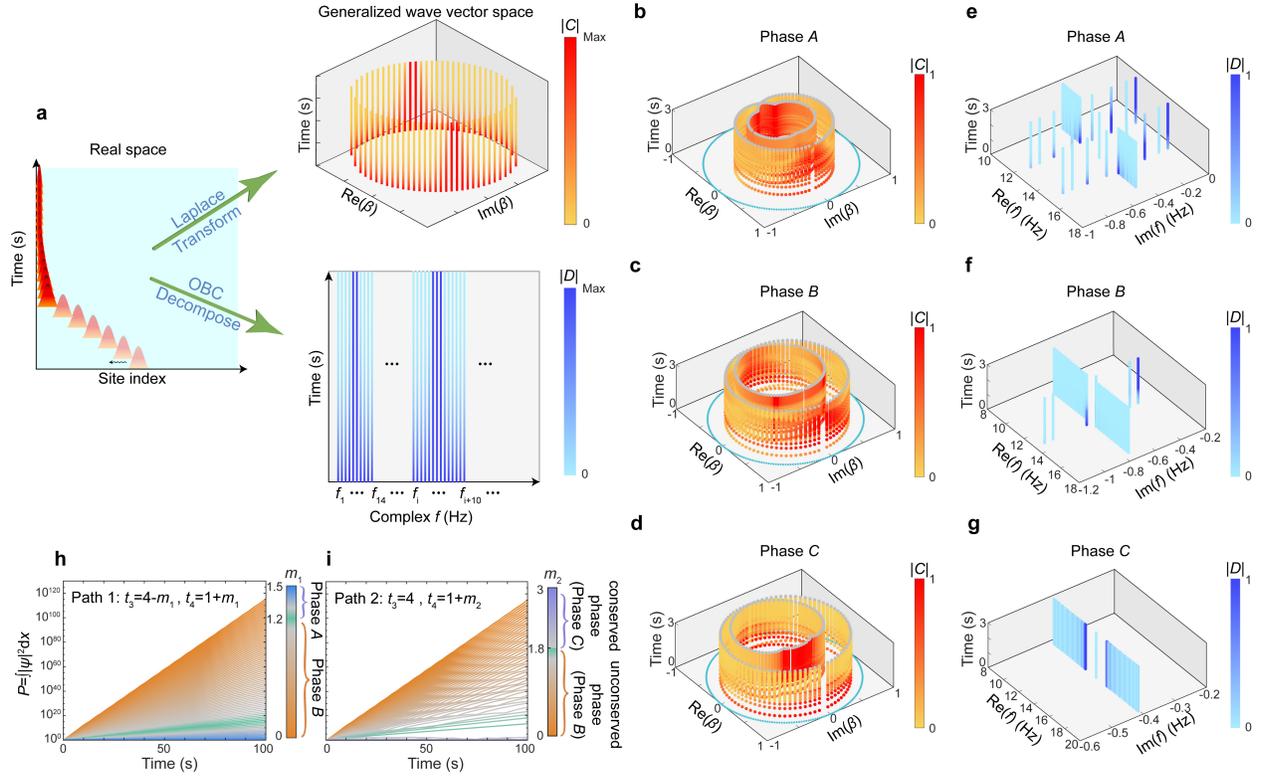

**Figure 5 | Understanding dynamics and phase transitions with GBZs and OBC eigenmodes. a**, Schematic of the Laplace transformation and the OBC eigenmodes decomposition of the measured dynamics. $C(t)$ and $D(t)$ denote the components resulting from the two transformations, respectively. **b-d**, The time dependence of $|C(t)|$ derived by Laplace transformations on the experimental data presented in Fig. 4. The results are mapped to the corresponding GBZs. Here, the data is normalized at each instant. The $\beta$ values within the GBZ have been determined through calculations of the eigenvalues across 160 sites. The shapes of the GBZs are further delineated by the grey markers. **e-g**, $|D(t)|$ derived by decomposing calculated time-dependent wavefunction $\Psi(t)$ using the experimental parameters using the OBC eigenmodes. The data is normalized at each instant. **h-i**, Time-evolution of the energy of the non-Hermitian system for two different paths of the parameter evolution in the $t_3$-$t_4$ plane when $t_1 = 1$ and $t_2 = 2$ (depicted in Fig. 3d). **h**, Path 1: $t_3 = 4 - m_1$, $t_4 = 1 + m_1$. **i**, Path 2: $t_3 = 4$, $t_4 = 1 + m_2$. $m_1$ and $m_2$ are represented by the color bar.

Finally, transitions between different dynamic phases can be studied by examining the total energy of the system[49] along two paths in the phase diagram. Path 1 goes from phase *B* to phase *A*, while path 2 goes from the same origin to phase *C* (see Fig. 3d). As shown in Fig. 5h, from phase *B* to phase *A* along path 1 the wave amplification is gradually reduced. We notice that the reduction process is smooth, reflecting that both phases *B* and *A* are energy unconserved dynamic phases. In contrast, the results in Fig. 5i show that the transition from phase *B* to phase *C* along path 2 is not smooth, revealing a real transition in the amplification behavior, which is consistent with the fact that phase *C* is a special non-Hermitian dynamic phase with energy conservation.

In summary, we demonstrate rich non-Hermitian skin dynamics in a 1D non-Hermitian systems with *GT* symmetry. The observed bulk unidirectional wave propagation, the tunable wave amplification, and the boundary wave trapping provide unconventional manipulation of wave dynamics that are useful for guiding, trapping, and directionally amplifying waves in a controllable and robust way (Discussions on the disorder effect are presented in Supplementary Sec. II) and thus are promising for applications. For instance, the unidirectional wave propagation and temporal amplification enable simultaneous signal boosting and isolation that force signals to propagate and amplify in a preferred direction and suppress the propagation in the reversed direction. Such properties are beneficial for applications such as detection, imaging, transmission, and other signal-processing technologies. Our discovery unveils the fundamental aspects and the pivotal role of symmetry in non-Hermitian dynamics, providing insights into the underlying physics and paving the way toward understanding non-equilibrium phases with symmetry enrichment[50]--- an unfolding research frontier to be explored.

## Methods

**Experiments.** The onsite orbital is realized by the rotational oscillator, which consists of a programmable electric motor and a metal rotational arm anchored by two springs. The rotation of the motor is controlled by the microcontroller, which can send signals to the motor. The unit cell of the mechanical lattice utilized to realize the hopping parameters in phase *A* is shown in Fig. 3f. The reciprocal hopping $t_1, t_2, t_4$ is realized using tensioned springs coloured orange, blue, and cyan. Each onsite motor is equipped with two layers of rotational arms. The springs of the oscillator and the hopping springs $t_1$ and $t_2$ are connected by the lower rotational arm with a length of 130 mm. The inter-chain hopping spring $t_4$ is connected by the upper rotational arm with a length of 93 mm. The observed dynamic skin effects in phases *A*, *B*, and *C* are located in the lower region of the Hermitian line, requiring a unidirectional hopping from site 1 (site 4) to site 2 (site 3). To achieve this, tailor-made circuitry is employed for nonreciprocal hopping. When a signal is sent to site 1, the real-time rotation angle $\theta_1$ ($\theta_4$) is measured and then a torque $\tau_2$ ($\tau_3$) is generated at site 2 (site 3) through the microcontroller, which is directly related to the angle at site 1 (site 4) ($\tau_2 = \alpha_{21}\theta_1, \tau_3 = \alpha_{34}\theta_4$). Here, $\alpha_{ij}$ represents a tunable coefficient. In cases where the loss needed to balance the amplification is increased (e.g., for systems in phase *B*), a passive motor is used as the additional attenuator to dissipate the energy (see Supplementary Information Sec. IV for more information).

In the experiments, each 1D system consists of 40 mechanical oscillators with hoppings and dampings designed according to the above approaches. The initial excitation is a constant rotation angle at an oscillator in the middle of the system. The data, i.e., the instantaneous rotational angles, were collected automatically using a sampling rate of 500 Hz.

**Laplace transformation to the GBZ.** Expanding from the wave vector $k$ in the Brillouin zone (BZ) to the generalized wave vector $\beta$ in the GBZ can be accomplished by employing the identity $\beta \equiv e^{iq}$, where $q \in \mathbb{C}$. Consequently, the Bloch Hamiltonian, as exemplified by Eq. (1), undergoes a transformation and is rewritten as the non-Bloch Hamiltonian

$$H(\beta) = \begin{pmatrix} 0 & t_4 & t_2 + t_1/\beta & 0 \\ t_3 & 0 & 0 & t_1 + t_2/\beta \\ t_2 + t_1\beta & 0 & 0 & t_3 \\ 0 & t_1 + t_2\beta & t_4 & 0 \end{pmatrix} \quad (M1)$$

Similarly, the final state under OBC $\psi(t, x)$ at any given time can be expressed as

$$|\psi(t, x)\rangle = U(t, t_0)|\psi(t_0, x)\rangle \quad (M2)$$

where $U(t, t_0)$ represents the time evolution operator $U(t, t_0) = e^{-iH_{OBC}t}$ and $\psi(t_0, x)$ denotes an initial state excited on a Hamiltonian $H_{OBC}$ under OBC. Subsequently, we can perform a Laplace transformation of $\psi(t, x)$ to acquire its representation $\Psi_j(t, \beta)$ in the generalized momentum space,

$$\Psi_j(t, \beta) = \int_{-\infty}^{\infty} \psi(t, x) \beta^{-x} dx \quad (M3)$$

Through the analysis of the final state $\Psi_j(t, \beta)$, one can discern the generalized wave vectors $\beta$ that exhibit substantial contributions during the evolution process. Next, we decompose the final state $\Psi_j(t, \beta)$ in generalized momentum space into the eigenstates of the non-Bloch Hamiltonian, obtaining the component coefficients $C_j(t, \beta)$,

$$|\Psi_j(t, \beta)\rangle = \sum_j C_j(t, \beta) |\varphi_{j,R}(\beta)\rangle, \quad (M4)$$

$$C_j(t, \beta) = \langle \varphi_{j,L}(\beta) | \Psi_j(t, \beta) \rangle, \quad (M5)$$

where $\varphi_{j,R}(\beta)$ and $\varphi_{j,L}(\beta)$ are the right and left eigenvectors of the non-Bloch Hamiltonian, respectively, and can be obtained using the following equation.

$$H(\beta) |\varphi_{j,R}(\beta)\rangle = E_j(\beta) |\varphi_{j,R}(\beta)\rangle, \quad (M6)$$

$$\langle \varphi_{j,L}(\beta)| H(\beta) = E_j(\beta) \langle \varphi_{j,L}(\beta)|. \quad (M7)$$

Additionally, $\varphi_{j,R}(\beta)$ and $\varphi_{j,L}(\beta)$ satisfy the biorthogonality condition,

$$\langle \varphi_{i,L}(\beta) | \varphi_{j,R}(\beta) \rangle = \delta_{i,j} \quad (M8)$$

**Non-Hermitian dynamics in the OBC eigenmodes.** At any given moment, the final state $\Psi(t)$ can be obtained by applying the time evolution operator $U \equiv e^{-iH_{OBC}t}$ to the initial state $\Psi(0)$.

$$\Psi(t) = e^{-iH_{OBC}t} \cdot \Psi(0) \quad (M9)$$

We can then decompose the final state $\Psi(t)$ by the eigenstates

$$\Psi(t) = \sum_j D_j(t) |\varphi_{j,R}\rangle \quad (M10)$$

This yields the coefficients $D_j(t)$

$$D_j(t) = \langle \varphi_{j,L} | \Psi(t) \rangle \tag{M11}$$

**Spatial symmetry of the system.** The system has the glide reflection symmetry $G$, beside the time-reversal symmetry $T$. However, there is no mirror or inversion symmetry. The model does have some interesting properties under these transformations, which are listed in the table below. We note that the unprimed and primed phases in the phase diagram (Figs. 2f, 3d) are related by the mirror transformation $x \to -x$. As a result, the phase diagram is symmetric with respect to the Hermitian line $t_3 = t_4$. From Table I, one can see that the direction of the dynamic NHSE in our system can be reversed either by switching $t_3$ and $t_4$, or by switching $t_1$ and $t_2$.

| Symbols | Operation | Effects |
|---|---|---|
| $M_x$ | $x \to -x$ | $t_3 \leftrightarrow t_4$ |
| $M_y$ | $y \to -y$ | $t_3 \leftrightarrow t_4, t_1 \leftrightarrow t_2$ |
| $P$ | $(x, y) \to (-x, -y)$ | $t_1 \leftrightarrow t_2$ |
| $G$ | $(x, y) \to (x + \frac{1}{2}, -y)$ | Invariant |

**Table I | Spatial transformations and their effects.** The first column lists the symbols of the spatial transformations. The second column gives the operation for each spatial transformation. The third column provides the effects of these spatial transformations on the parameters of our model. Here, we set the origin of the coordinates at the unit cell center.

## Acknowledgments

This work was supported by the National Key R&D Program of China (2022YFA1404400). J.H.J also thanks the supports from the National Natural Science Foundation of China (Grant Nos. 12125504 and 12074281), and the Priority Academic Program Development (PAPD) of Jiangsu Higher Education Institutions. G.M. acknowledges the supports from the Hong Kong Research Grants Council (#12301822 #RFS2223-2S01).

## Author contributions

J.H.J. and G.M. guided the research. L.W.W., Z.K.L., and J.H.J. proposed and established the theory. Z.L., X.W., and G.M. designed and performed the experiments. J.H.J. and G.M. wrote the main text. L.W.W., Z.L., and G.M. wrote the Supplementary Information. All authors were involved in the discussions of the results.

## Competing Interests

The authors declare that they have no competing financial interests.

## Data availability

All data are available in the manuscript and the Supplementary Information. Additional information is available from the corresponding authors through reasonable request.

## Code availability

Reasonable requests for the computation details can be addressed to the corresponding authors.